\def\fs{\footnotesize}

 \def\mso{\,\mathrm{M}_\odot}

 \def\kms{\, {\rm km}\, {\rm s}^{-1}}

 \def\simle{\mathrel{\hbox{\rlap{\hbox{\lower4pt\hbox{$\sim$}}}\hbox{$<$}}}}
 \def\simgr{\mathrel{\hbox{\rlap{\hbox{\lower4pt\hbox{$\sim$}}}\hbox{$>$}}}}

\def\fs{\scriptsize}

\documentclass{iaus}
\usepackage{graphicx, natbib}
\bibpunct{(}{)}{;}{a}{}{,}

\title[Rotation and massive close binary evolution] %% give here short title %%
{Rotation and massive close binary evolution}

\author[N. Langer, M. Cantiello, S.-C. Yoon, et al.]   %% give here short author list %%
{Norbert Langer$^1$, Matteo Cantiello$^1$, Sung-Chul Yoon$^2$, Ian Hunter$^3$, Ines Brott$^1$, 
Danny Lennon$^4$, Selma de Mink$^1$ \& Marcel Verheijdt$^1$}

\affiliation{$^1$Astronomical Institute, Utrecht University, The Netherlands \\[\affilskip]
$^2$University of California, Santa Cruz, USA \\[\affilskip]
$^3$The Queen's University of Belfast, Northern Ireland, UK \\[\affilskip]
$^4$Isaac Newton Group, Santa Cruz de La Palma, Canary Islands, Spain}

\pubyear{2008}
\volume{250}  %% insert here IAU Symposium No.
\pagerange{0--0}
\jname{Massive Stars as Cosmic Engines}
\editors{F. Bresolin, P. Crowther \& J. Puls, eds.}
\begin{document}

\maketitle

\begin{abstract}
We review the role of rotation in massive close binary systems.
Rotation has been advocated as an essential ingredient in massive single
star models. However, rotation clearly is most important in massive
binaries where one star accretes matter from a close companion,
as the resulting spin-up drives the accretor towards critical rotation. 
Here, we explore our understanding of this process, and its observable
consequences. When accounting for these consequences, the question remains
whether rotational effects in massive single stars are still needed
to explain the observations. 

\end{abstract}

\firstsection % if your document starts with a section,

\section{Why look at rotation?}

Rotation has been identified as an important physics ingredient which needs 
to be considered to understand the evolution of massive star
(e.g., Heger, Langer \& Woosley 2000; Meynet \& Maeder 2000). 
It is thought to gives rise to physical effects inside stars which
cause observable quantities to change, and may even
radically alter the evolutionary path of the stars. 
A drastic example is the occurrence of chemically homogeneous evolution,
which may provide a progenitor path towards long gamma-ray bursts
(Yoon \& Langer 2005, Yoon et al. 2006, Woosley \& Heger 2006).

One of the most relevant prediction of massive star models with rotation is
that rotationally triggered internal transport processes are capable to 
bring nuclear processed material, most notably nitrogen, 
from the convective core of massive main sequence stars into their radiative envelope.  
For fast enough rotation, fresh nitrogen thus appears at the surface of the
star, and becomes continuously more enriched as function of time during core
hydrogen burning. 

Numerous incidental evidences have been collected from observations which are
in support of this picture (cf. Maeder \& Meynet 2000, and references therein).
However, while many observations refer to stars in their post-main sequence
stages, abundance analyses of main sequence stars have mostly been restricted
to apparent slow rotators and to rather small groups of stars.
A major step forward in comparing massive star models and observations is provided
by the FLAMES Survey of Massive Stars, which encompassed many hundred O and early B 
main sequence stars in the Galaxy and the Magellanic Clouds (Evans et al. 2005).
The B~star sample of this survey was analyzed in a way which allowed for the first
time to obtain quantitative constrains on the nitrogen enhancement also in a large number of
rapid rotators (Hunter et al. 2008, Fig.~1). Surprisingly, Hunter et al. could not
unambiguously conclude that the effects of rotation are observed as expected. 
They rather found two groups of rapid rotators, one with significant enrichment
and one without (the latter being dubbed Group~1 in Fig.~1). 
Both groups contain stars which are close to core hydrogen
exhaustion, as indicated by their low surface gravity. 

\begin{figure}[t]
\begin{center}
\includegraphics[width=0.98\textwidth]{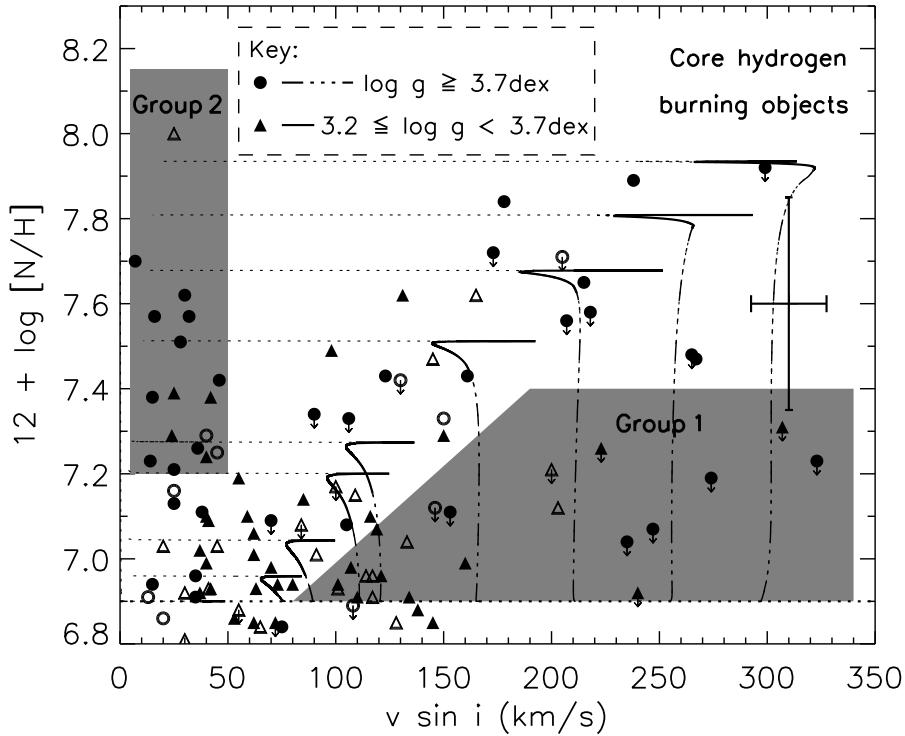}
\caption{Nitrogen abundance (12 + log [N/H]) against the projected
rotational velocity ($v\sin i$) for core hydrogen burning objects in two LMC
fields centered on N11 and NGC2004, according to Hunter et al. (2008).
Open symbols: radial velocity variables; downward arrows: abundance upper limits;
lowest dotted line: LMC baseline nitrogen abundance.
The mean uncertainty in the nitrogen abundance is 0.25\,dex while that in $v\sin i$ is 10\%.
The bulk of the stars occupy a region at low
$v\sin i$ and show little or modest nitrogen enrichment.
The stellar evolution tracks are computed for an initial mass of 13$\mso$
corresponding to the average mass of the sample stars,
and their rotational velocity has been multiplied by $\pi$/4 to account for random
inclinations.
The surface gravity is used as indicator of the evolutionary status and the objects
(see legend) and tracks have been split to
indicate younger and older core hydrogen burning stars, respectively.
Gray shading highlights two groups of stars which remain unexplained
by the stellar evolution tracks. }
\end{center}
\end{figure}

Hunter et al. (2008) gave two possible ways of interpretation. The one which saves
the current picture of rotational mixing is that the enriched fast rotators in
the FLAMES sample are indeed single stars, while the non-enriched fast rotators
have a peculiar binary history. An observing campaign is underway to test the hypothesis
that these latter stars are indeed all binaries --- for which in the current FLAMES
data there is no clear evidence. Alternatively, the results of Hunter et al. (2008)
could imply that rotational mixing is not efficient, and that the enriched fast rotators
are all spun-up accretion stars in binaries. The worry that the latter might be true
is strengthened by the finding of yet another discrete group of massive main sequence
stars by Hunter et al. (2008; named Group~2 in Fig.~1), 
namely intrinsically slowly rotating stars with a
strong nitrogen enhancement. While this group of stars clearly needs an alternative
explanation, it appears likely that previous reports of nitrogen enrichment in 
massive main sequence stars which served as support for rotational mixing picked up
stars comparable to those in Group~2, as they were limited to low projected rotational velocities. 

In the following, we discuss which possibilities are supported by current models of massive close
binaries. As mentioned above, it appears impossible to understand
the nitrogen pattern in fast rotators without invoking close binaries. 
One may actually wonder whether even {\em all} fast rotators could be produced
by close binary effects.

\section{Required physics}

Massive close binary evolution is modeled by various groups (e.g., 
Podsiadlowski et al. 1992, Wellstein \& Langer 1999, Belczynski et al. 2002,
Vanbeveren et al. 2007, Vazquez et al. 2007). However, in order to
predict the surface nitrogen abundances and the rotational velocities
of binary components, a rather large amount of physical effects needs to
be considered in binary evolution models.

It is desirable to include, in such binary models, the physics of rotation as
it is currently used in models of rotating massive single stars. The reason is
that in close binaries, even rather small amounts of matter transferred during
Roche-lobe overflow spins up the mass gainer to extreme
rotation rates (Packet 1981). Therefore, {\em if} rotational mixing is 
real, it might have the strongest effects in binary systems. 

\begin{figure}[t]
\begin{center}
\includegraphics[angle=270,width=0.98\textwidth]{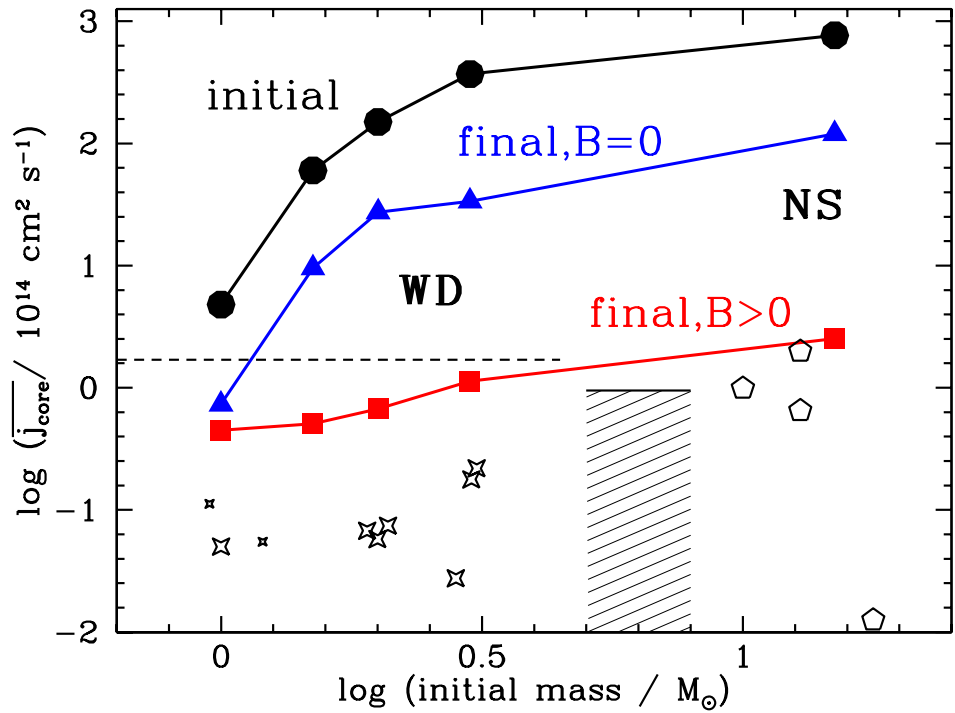}
 \caption{
         Average initial (upper line) and final core specific angular momentum
         of 1...3$\mso$ stars according to Suijs et al. (2008),
         and 15$\mso$ stars according to Heger et al. (2005).
        Filled triangles corresponds to the final
        models of non-magnetic sequences, and filled squares
        to the final models of magnetic sequences. 
        The dashed horizontal line indicates the spectroscopic upper limit  
        on the white dwarf spins obtained by Berger et al. (2005).
        Star symbols represent astroseismic measurements from ZZ Ceti stars
        (Bradley 1998, 2001; Dolez 2006; Handler 2001, Handler et al. 2002, Kepler et al. 1995,
        Kleinmann et al. 1998, Winget et al. 1994),
        where smaller symbols correspond to less certain measurements.
        The green hatched area is populated by magnetic white dwarfs
        (Ferrario \& Wickramasinghe 2005; Brinkworth et al. 2007).
        The three black open pentagons correspond to the youngest Galactic
        neutron stars (Heger et al. 2005). The lowest pentagon is
        thought to roughly correspond to magnetars (Camilo et al. 2007).}
\end{center}
\end{figure}

Angular momentum transport by internal magnetic fields is one ingredient in single 
star models which appears to be indispensable as well. Heger et al. (2005) showed
that without this effect, young neutron stars are predicted to spin too rapidly.
Suijs et al. (2008) showed that magnetic transport is also required to prevent
too fast rotation in white dwarfs (Fig.~2). 

For binary evolution models, there are two more pieces of physics which need to be
included, which both relate to angular momentum exchange between the components of
a binary system. Mass transfer within the Roche approximation is commonly applied in 
binary evolution calculations, but the corresponding angular momentum transfer is
mostly neglected. The latter is crucial to model the spin-up of the accretion star,
and thus to represent the most rapidly rotating stars at all. Spin-orbit coupling
through tides is the other unmissable ingredient in close binary models, as it can
lead to significant spin-down (mostly!) or spin-up (rarely) in massive binaries
with periods below 10...20~d. 

Wellstein (2001) and Petrovic et al. (2005, 2005a; see also Detmers et al. 2008) 
have produced a binary code which includes all the required physics.
While only few binary evolution models have been computed with this code so far,
these models may help to answer a few of the questions raised above.
Some of their properties are discussed below.

\section{Luminosity and effective temperature}

\begin{figure}[]
\begin{center}
\includegraphics[width=0.78\textwidth]{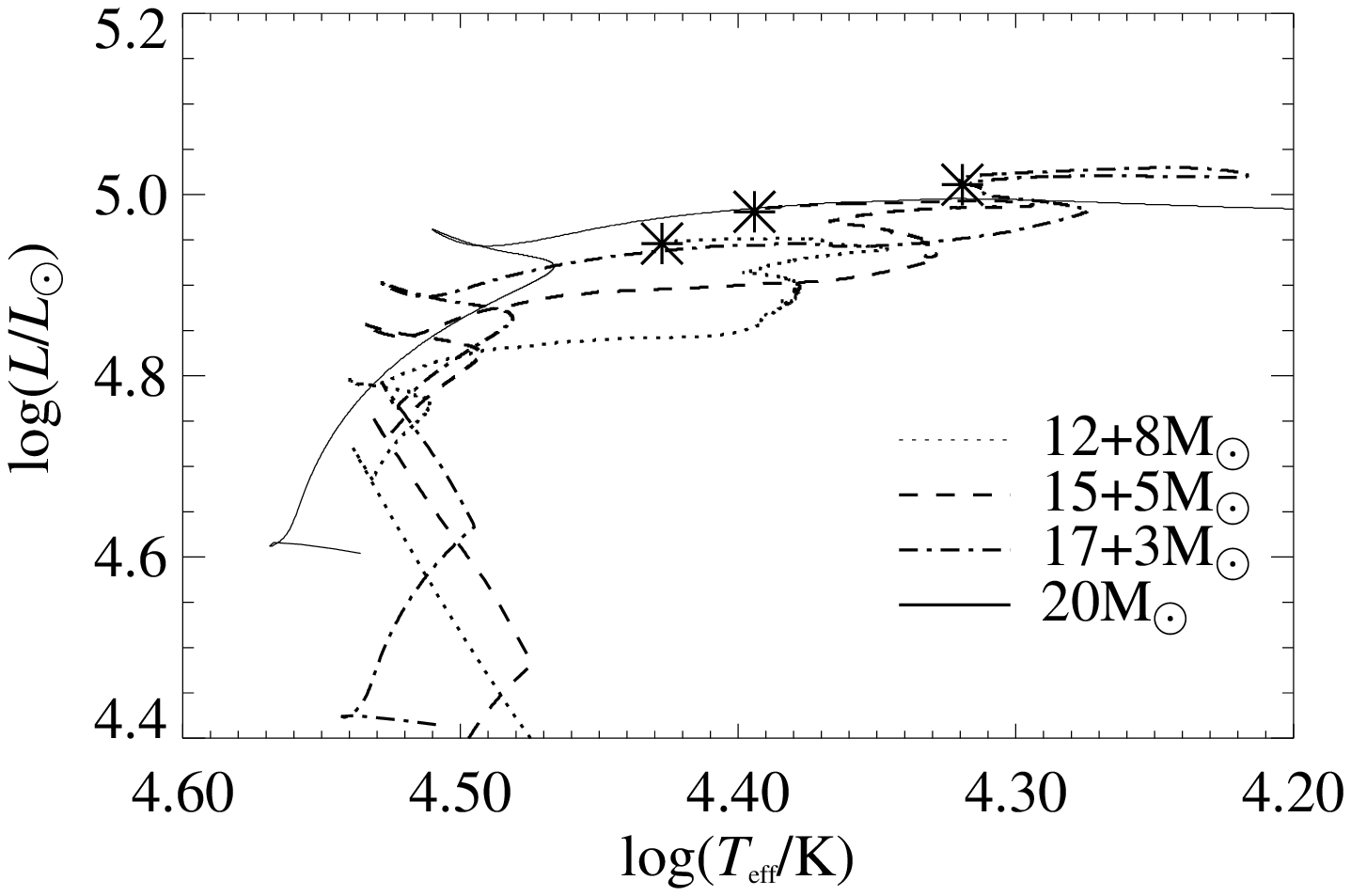}
\caption{Tracks in the HR diagram for stars with a total mass of 20$\mso$ 
after accretion at a central helium mass fraction of $Y=0.7$,
starting at 12$\mso$, 15$\mso$, and 17$\mso$ (see legend), compared to
the track of a 20$\mso$ single star (Braun \& Langer 1995). While the single star evolves
to the red supergiant stage immediately after core hydrogen exhaustion,
the accreting stars, which do not rejuvenate, remain blue supergiants
throughout core helium burning. Their pre-supernova position is indicated be
an asterisk. }
\end{center}
%\end{figure}
%\begin{figure}[]
\begin{center}
\includegraphics[width=0.78\textwidth]{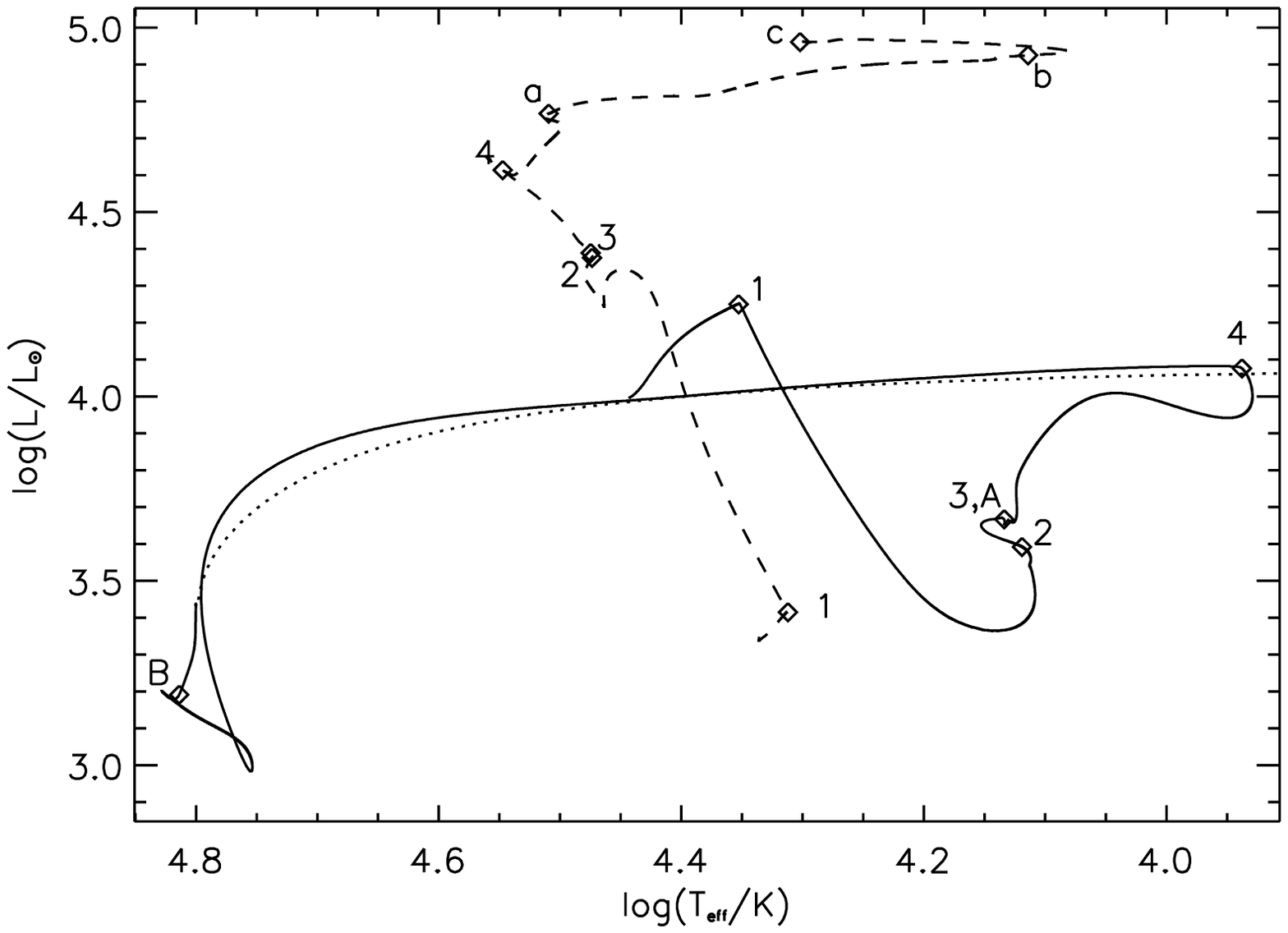}
\caption{Evolutionary tracks of the primary
  (solid and dotted line) and secondary star (dashed line) of the 
  case~A binary system No.~31 of Wellstein et al. (2001; initial masses are $12\mso$ and
  $7.5\mso$, the initial period is 2.5$\,$d)
  in the HR~diagram.  Beginning and end of the mass transfer
  phases are marked with numbers; 1: begin of Case~A, 2: end of
  Case~A, 3: begin of Case~AB, 4: end of Case~AB.  The labels A/a
  designate the end of central hydrogen burning of the
  primary/secondary, B/b the end of central helium burning of the
  primary/secondary, and c the point of the supernova explosion of the
  secondary. In this system, the secondary star ends its evolution
  first. The time of its supernova explosion marks the end of the
  solid line in the track of the primary. The further evolution of the
  primary is shown as dotted line. During this phase, it is treated as
  a single star since the system is likely broken up due to the
  secondary's explosion.}
\end{center}
\end{figure}

We want to briefly discuss two important effects of binarity on the distribution
of stars in the HR diagram, in particular concerning the mass gainer, which
will be the more prominent star of the two after a mass transfer event.

Fig.~3 shows that mass gainers become more luminous the more mass they gain.
However, if the accreted amount of matter is large, and if it occurs late
enough during the core hydrogen burning evolution of the mass gainer,
its rejuvenation, i.e. in particular the growth of its convective
core to adapt to the increased stellar mass, might be avoided (Braun 
\& Langer 1995). In this case, core helium burning may take place 
very close to the main sequence, i.e., helium burning stars may be mistaken
for main sequence stars. In this respect, we point out that Hunter et al. (2008) 
interpreted the sharp drop of the projected rotational as function of surface 
gravity as signaling the cool end of the main sequence band.  

The increased luminosity of the mass gainers, which is also clearly visible in
the evolutionary tracks shown in Fig.~4, may cause them to appear as blue stragglers
in samples of stars with similar age (cf., Pols \& Marinus 1994).
If then a sample of stars is defined through a visual magnitude cut-off,
which might favor evolved main sequence stars near the turn-off,
it is conceivable that mass gainers constitute a significant fraction
of the whole sample.

\section{Distribution of rotational velocities}

In order to model the distribution of stars in Fig.~1, 
one requires a distribution function for the initial rotational
velocity of single stars (IRF). However, what can be measured is
only the present-day distribution of rotational velocities (PRF). 
Fig.~5 gives an example derived within the FLAMES survey, which is
the PRF of the O~stars in an SMC field centered on NGC~346 (Mokiem et al. 2006).
The Magellanic Clouds are good study grounds for this, since at least spin-down
of single stars by radiation driven winds can be neglected for all except
the very most massive main sequence stars. Worrisome about Fig.~5 is that
all but the fastest three rotators have rotation rates of less than 
a quarter of critical rotation, while two of the three fast rotators appear to
be runaway stars. We thus ask the question: could {\em all} rapid rotators be
binary products?

\begin{figure}[]
\begin{center}
\includegraphics[width=0.98\textwidth]{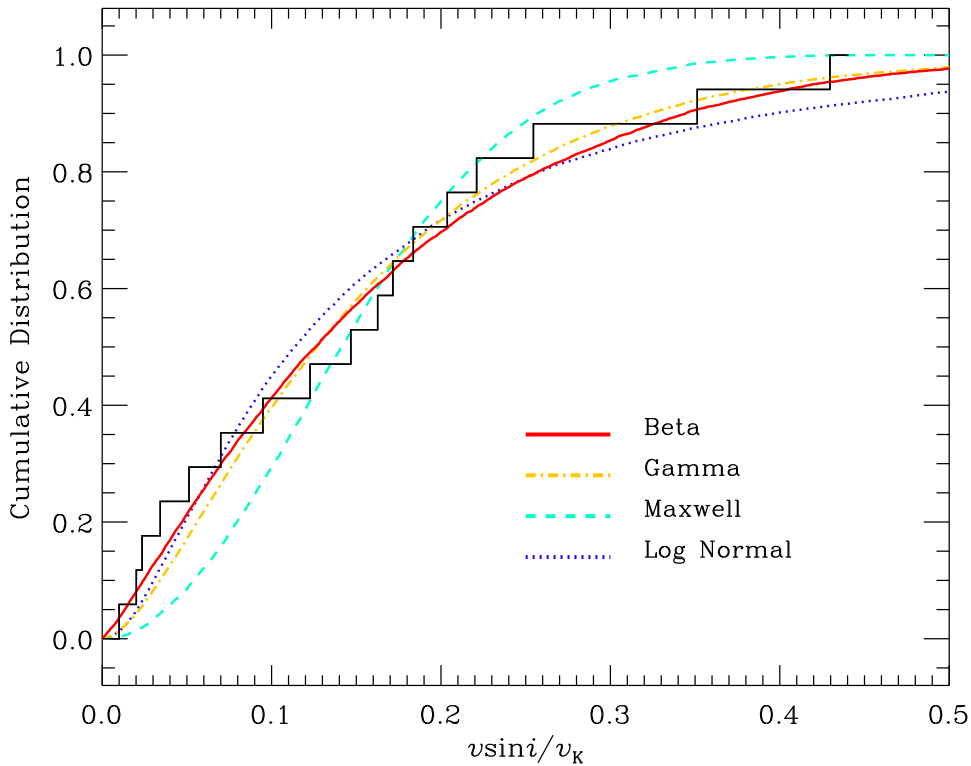}
\caption{Cumulative distribution of the fraction of the Keplerian value
of the observed rotational velocity (i.e., $v\sin i$) of unevolved young stars in NGC~346 in the
Small Magellanic Cloud, according to Yoon et al. (2006).
The data (step function) is from Mokiem et al. (2006). The dotted-dashed, solid, and dashed lines
are the best fits of synthesized distribution functions using three different distribution laws:
beta, gamma and Maxwellian, respectively. Here it is assumed that the
stellar rotation axes are randomly oriented.}
\end{center}
\end{figure}

\begin{figure}[t]
\begin{center}
\includegraphics[width=0.98\textwidth]{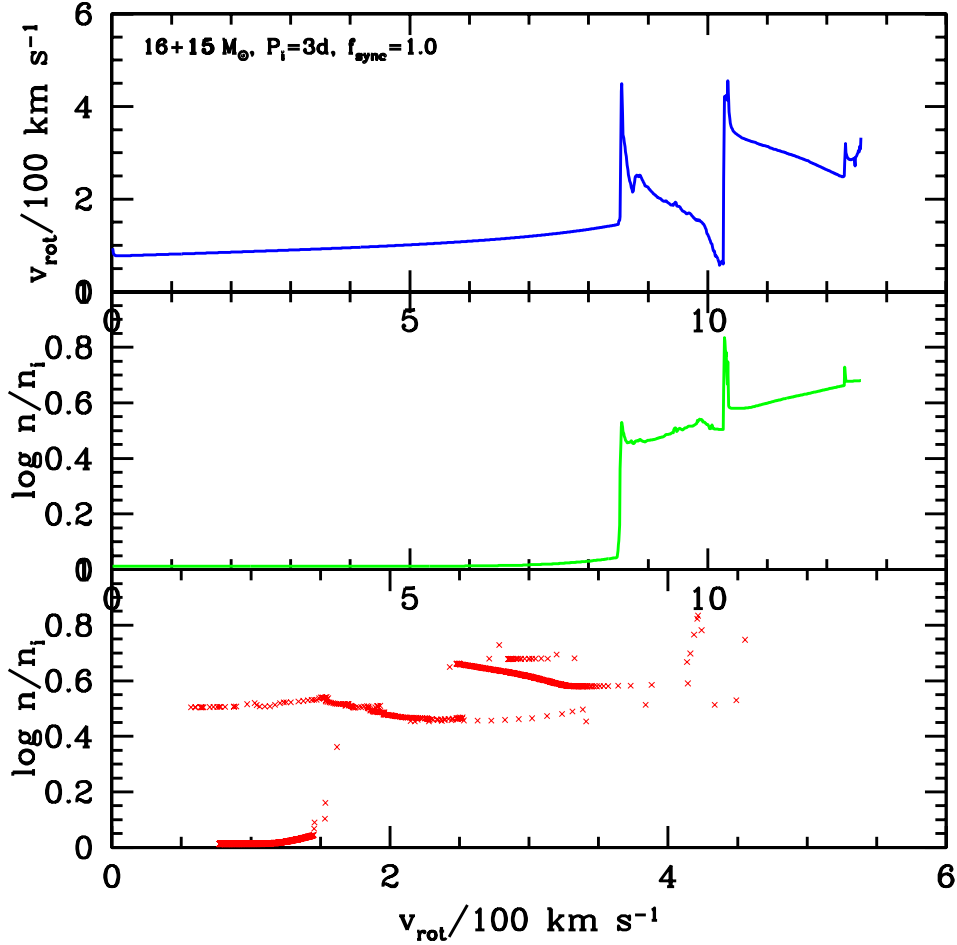}
\caption{Equatorial rotational velocity (upper panel) and
surface nitrogen mass fraction relative to the initial value
(middle panel) as function of time, for a the mass gainer in a 
solar metallicity
$16\mso + 15\mso$ binary with an initial period of 3~days.
Time is given in Myr for the upper two panels. 
The computations include the physics of rotation for both
components as in Heger et al. (2000), and Spin-Orbit coupling
as in Detmers et al. (2008) with the nominal coupling parameter
$f_{\rm sync}=1$, and rotationally enhanced stellar 
wind mass loss (Langer 1998). Internal magnetic fields are not
included. The bottom panel shows the evolution of the mass gainer
in the nitrogen enhancement versus rotational velocity diagram,
where each data point represents a duration of 20\,000\,yr.
The spin-down of the star after the first accretion event
($t=8.5...10\,$Myr) is mostly due to tidal effects. 
This example shows that massive close binaries can produce
rotating nitrogen rich stars which are rapidly rotating,
but also such which are slowly rotating.}
\end{center}
\end{figure}

\begin{figure}[t]
\begin{center}
\includegraphics[width=0.98\textwidth]{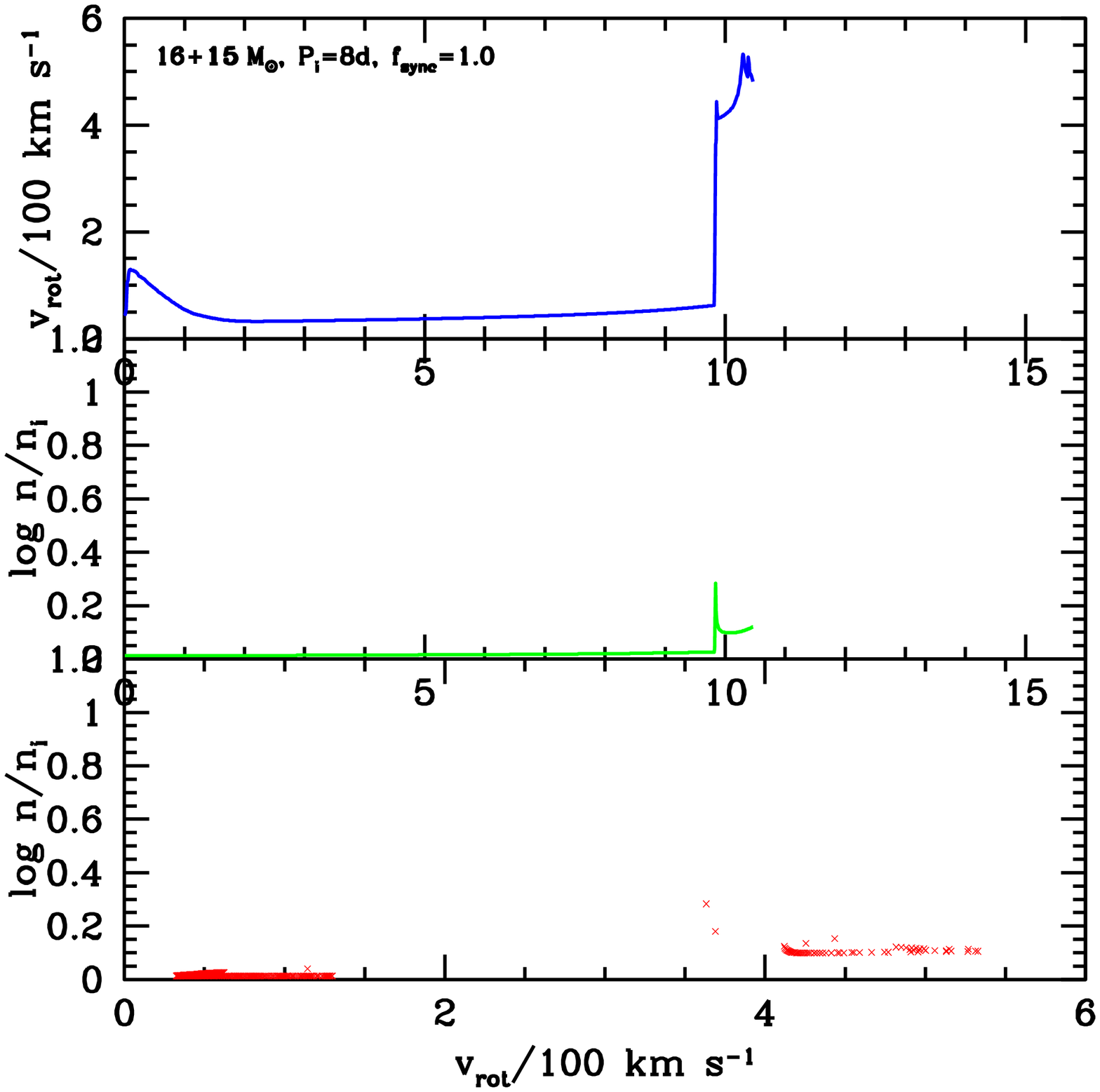}
\caption{As Fig.~6, but for a 
$16\mso + 15\mso$ binary with an initial period of 8~days,
which makes it a Case~B system.
Due to the larger initial period, the mass transfer is very
non-conservative. Enough mass is accreted to spin-up the mass
gainer, but not enough to create a large nitrogen surface 
enhancement. 
This example shows that massive close binaries can produce
rapidly rotating, evolved main sequence stars which are
{\em not} strongly nitrogen-enriched. Note that the calculations
stops about 1\,Myr after the mass transfer due to numerical
difficulties. Possibly, the mass gainer would become 
nitrogen-enhanced later on due to rotational mixing.}
\end{center}
\end{figure}

The top panels of Figs.~6 and~7 show the evolution of the rotational
velocity of mass gainers for two different binary evolution models
(Wellstein 2001). The mass gainer will be the dominantly visible star after  
the first accretion event, and the mass loser may be hard to notice at all
or even be ejected through its supernova explosion. These figures show that
very rapid rotators are produced, which are long-lived main sequence
stars. Due to the blue straggler effect, they may dominate certain parts
of the HR-diagram (cf. Section 3). 

\begin{figure}[]
\begin{center}
\includegraphics[width=0.98\textwidth]{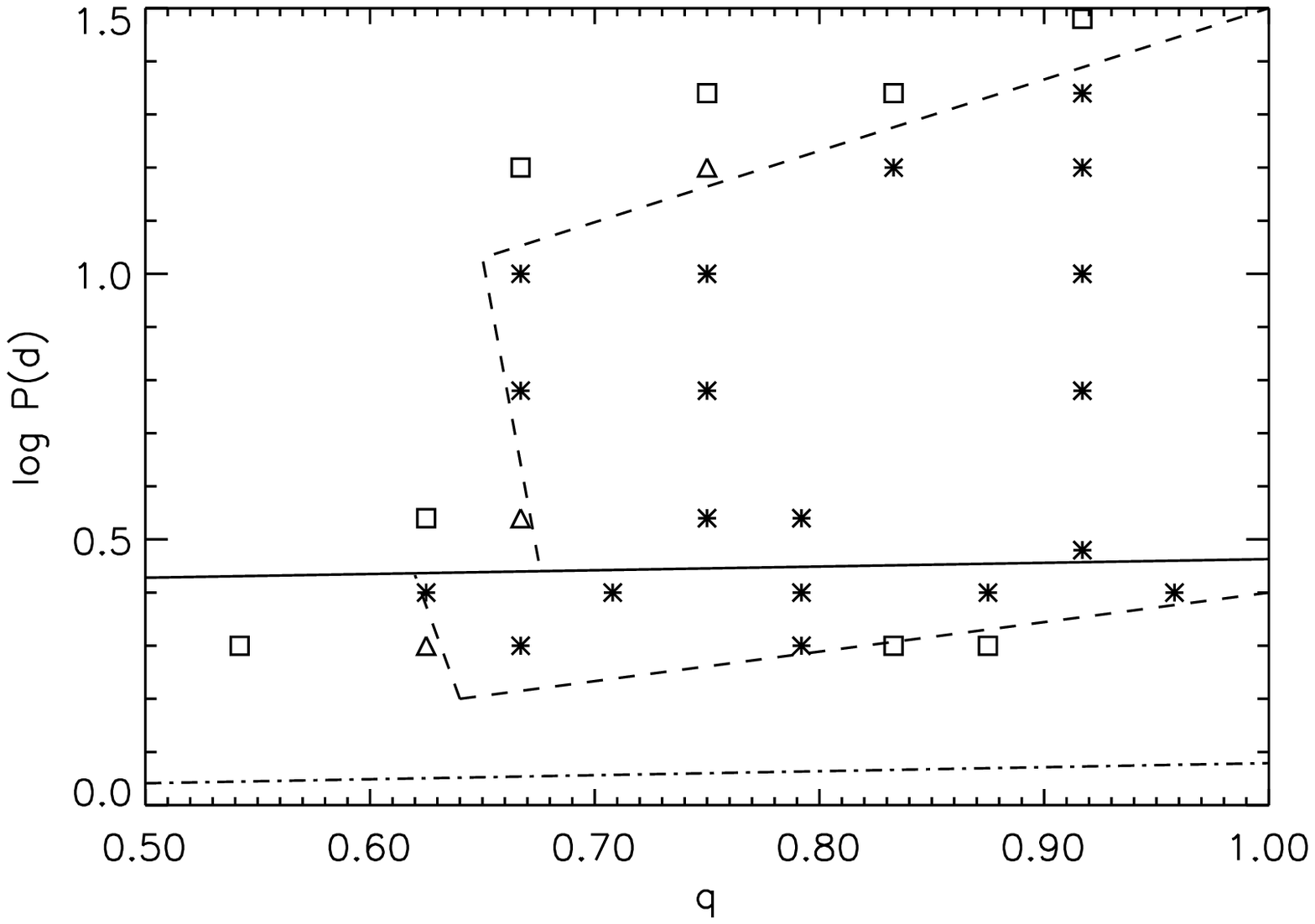}
\caption{Distribution of all computed binaries with 12$\mso$ primaries 
  of Wellstein et al. (2001) in the initial period versus initial mass ratio
  diagram. Asterisks mark contact-free systems, while
  squares mark systems which evolve into contact. Systems marked with
  triangles are borderline cases, i.e., they evolve into a short
  contact phase but the secondary radius never exceeds its Roche
  radius by more than a factor~1.5. The solid line separates Case~A
  (below) and Case~B systems. All case~A systems for this primary mass
  have a reverse supernova order. The dashed lines indicate the
  boundary between contact-free and contact evolution. The
  dashed-dotted line is defined by the condition that the primary
  fills its Roche lobe already on the zero age main sequence.}
\end{center}
\end{figure}

Fig.~8 shows the range of contact-free evolution in the initial period 
versus initial mass ratio diagram for primary star masses of relevance here.
Mass transfer and spin-up is expected everywhere within the contact-free
regime (Wellstein et al. 2001). However, in most regions outside of this, 
both stars in the binary are expected to merge as a result of their interaction
(see also Podsiadlowski et al. 1992). While the details of the merger process
are difficult to predict, the merger product will be an extreme rotator due to
the enormous surplus of angular momentum. Merger stars will only be observed as
single stars.  

We see that due to mass transfer and merging, a stellar population with few
or no rapid rotators initially may build up a certain number of rapidly
rotating core hydrogen burning stars. Some of them may slow down again due to
their close companion (cf. Fig.~6, top panel), but many will not.
It will be an important task for the near future to put quantitative limits
on the relative number of those stars in stellar populations.

\section{Nitrogen enrichment}

\begin{figure}[]
\begin{center}
\includegraphics[angle=270,width=0.98\textwidth]{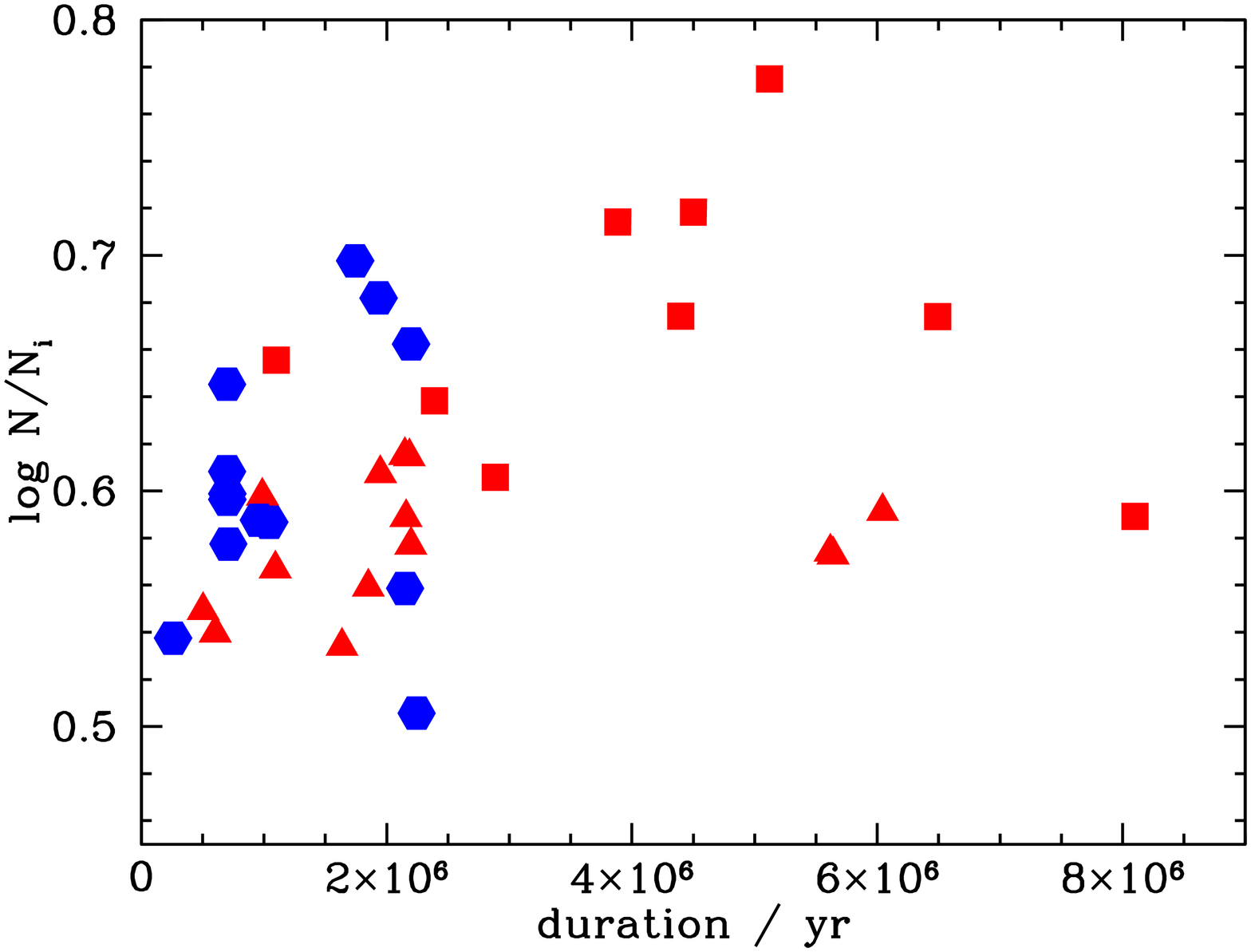}
\caption{Surface abundance of nitrogen 
      relative to its initial abundance
      for the solar metallicity binary models
      of Wellstein et al. (2001),
      as function of the duration of the post-mass transfer phases.
      The models are conservative, have 12$\mso$ and 16$\mso$ primaries, 
      and initial mass ratios larger than 0.5. 
      Case~A mass gainers during slow Case A mass transfer
      (Algol-type systems) are marked by triangles.
      The same stars appear as squares after Case~AB mass transfer.
      Post Case~B mass gainers are marked by hexagons.}
\end{center}
\end{figure}

\begin{figure}[]
\begin{center}
\includegraphics[width=0.98\textwidth]{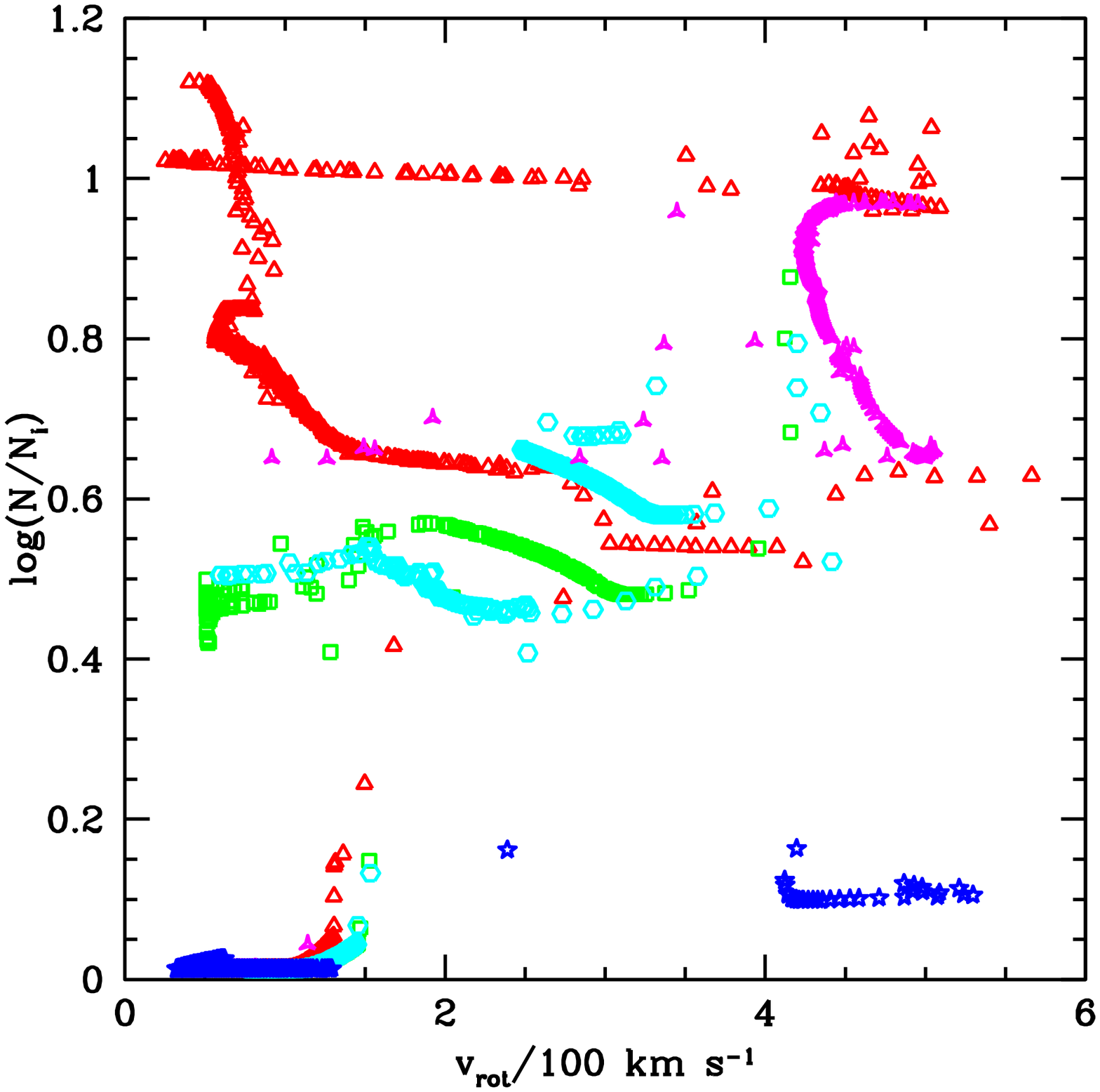}
\caption{Nitrogen enhancement versus rotational velocity diagram,
for the mass gainers of 5 computed binary systems, 
where each data point represents a duration of 20\,000\,yr.
All but one system start out with 16$\mso$+15$\mso$.
Open hexagons are used for the Case~A system shown in Fig.~6,
while open squares show the analogous system with a 10 time
stronger tidal force parameter. Three-spiked stars designate 
a Case~B system with an initial period of 6~d, while five-spiked stars
show the 8~d system from Fig.~7. Triangles mark a 10$\mso$+9$\mso$
Case~A system with an initial period of 2.15~d. Only the core-hydrogen burning
stage of the mass gainers is shown. As in Fig.~9, it can be seen that 
the enrichment by mass transfer alone goes up to about $\log N/N_{\rm i} \simeq 0.6$,
with larger enrichments predominantly produced by rotational mixing.   
 }
\end{center}
\end{figure}

The surface nitrogen abundance of mass gainers in close binary systems 
can be affected in two major ways. First of all, the matter which is accreted from the
companion is often nitrogen-rich, as it stems from deep layers of the
initial primary. While the transferred matter may be very nitrogen-rich,
with values close to CNO-equilibrium being reached toward the end of the mass
transfer, the enrichment on the mass gainer remains limited since thermohaline
mixing dilutes the accreted matter with its whole envelope.
Figure~9 gives an idea of the maximum obtainable enrichment, which, in the mass range
considered here, amounts typically to 0.6~dex, where some Case~A systems can 
produce as much as 0.8~dex. Non-conservative systems, where much of the overflowing
matter is ejected from the binary system, are expected to obtain substantially 
smaller enrichments.

Secondly, nitrogen can be enhanced in mass gainers due to rotational mixing.
Since these stars are amongst the most rapidly rotating main sequence stars,
rotational mixing in these stars might be substantial.
In fact the second panel in Fig.~6 shows that, while accretion has raised
the nitrogen surface mass fraction by $\sim$0.4~dex, it furtheron slowly increases
to a total enrichement of almost $\sim$0.7~dex. 
Fig.~10 shows a more dramatic example: the same binary
system as shown in Fig.~6, but with an initial period of 6~d instead of 3~d,
accelerates its mass gainer to a rotational velocity of almost 500$\kms$,
which gives an extra nitrogen enrichment of more than a factor two, to 
about 1~dex in total. At low metallicity, Cantiello et al.
(2007) has found that the mass gainer of a 16$\mso$+15$\mso$ system with 
an initial period of 5~d evolves chemically homogeneously after the Case~B
mass transfer event.

So we see that, quite naturally, close binaries produce
rapidly rotating nitrogen-rich stars. This remains true if rotational mixing is
completely neglected, where the nitrogen enhancement would just be a factor of
2...3. With rotational mixing included, a factor of~10 can be reached. 
 
But Fig.~10 also shows that binaries can indeed produce rapid rotators with
little enrichment (cf. lower two panels in Fig.~7). Note that the highly
non-conservative evolution which is required to obtain this may be largely
underrepresented in Fig.~10, as the corresponding binary evolution models
are numerically difficult. 

Finally, as the post-mass transfer period in very close systems can lead to
rapid tidal spin-down (e.g., in the 2.15~d binary shown in Fig. 10), close binaries
may also produce slowly rotating nitrogen-rich main sequence stars. 

\section{Conclusions}

We have shown above that an effort is needed to compare the predictions of
stellar evolution theory with the new results derived from the FLAMES Survey
of Massive Stars (cf. Fig.~1). In particular, binary evolution models 
which include rotational mixing, mass and angular momentum transfer,
and tidal interaction are required, of which only few exist today
(cf., Fig.~10). 

Fig.~10 can not be directly compared with Fig.~1. It does not contain a clean
population study, and can therefore only indicate which parts of the diagram
might be populated by binary systems, but not how many stars one might expect
in those parts. Nevertheless, it appears rather striking that the few models
shown in Fig.~10 can populate each part of the diagram in which Fig.~1 shows
a significant density of stars. In fact, one may ask the question whether 
single stars are needed at all to understand Fig.~1, except for the slowly
rotating non-enriched stars. 

This question ties in with the one asked in Sect.~4: Could all rapid rotators
be spun-up mass gainers and merger products? While it seems difficult to 
answer this question currently, the following statement seems secure:
The FLAMES result shows that it is impossible to understand the 
nitrogen enhancement in massive stars without considering binary evolution.
Whether the same statement can be made for single stars is questioned 
by the results shown above. 
 
\acknowledgements{\fs SCY was supported by the DOE Program
for Scientific Discovery through Advanced Computing
and NASA.
 }

\end{document}